\documentclass{aa}  

\usepackage{graphicx}
\usepackage{txfonts}
\usepackage{hyperref}
\bibpunct{(}{)}{;}{a}{}{,} 

\begin{document} 

\title{Yet another star in the Albireo system}
\subtitle{The discovery of Albireo Ad}

\author{D. Jack \inst{1} \and  K.-P. Schr\"oder \inst{1,2} \and M. Mittag \inst{3} \and U. Bastian \inst{4}}
\institute{Departamento de Astronom\'ia, Universidad de Guanajuato,
Callej\'on de Jalisco S/N, 36023 Guanajuato, GTO, Mexico\\
              \email{dennis.jack@ugto.mx}
               \and Sterrewacht Leiden, Universiteit Leiden, Niels Bohrweg 2, NL-2333 CA Leiden, The Netherlands
               \and Hamburger Sternwarte, Universit\"at Hamburg, Gojenbergsweg 112, 21029 Hamburg, Germany
               \and Zentrum für Astronomie (Center for Astronomy), Heidelberg University, M\"onchhofstr. 14, D-69120 Heidelberg, Germany}
\date{Received XXX; accepted XXX}

\abstract
{Albireo is a well-known bright visual double star. It is still unclear 
if the components A and B form a gravitationally bound system.
The component Albireo~A is itself a binary star. 
The orbital parameters of the Albireo~Aa, Ac system have been determined only recently.
Thus, Albireo is still of interest for current research.}
{We aim to present evidence for the detection of a new member in the Albireo system. 
Furthermore, we aim to determine the 
orbital parameters and to find further conclusions for the Albireo system.}
{We used spectroscopic observations of Albireo~A obtained with the TIGRE telescope
and determined the radial velocities during a period of over three years.
We analyzed the radial velocity curve with RadVel to determine the orbital parameters. 
In addition, we determined the stellar parameters of Albireo~Aa with \texttt{iSpec}.}
{We found clear evidence for yet another star in the Albireo system 
orbiting Albireo~Aa with a period of about $P=371$~days. Several alternative
explanations for the periodic radial velocity signal could be discarded.
The new companion Albireo~Ad is a low mass star of about 0.085~$M_\odot$.}
{We conclude that Albireo is a hierarchical multiple star system and remains an interesting object for future observations and studies.}


\maketitle

%
%
\section{Introduction}

Albireo ($\beta$~Cyg) is a very bright double star easily visible for small-sized telescopes,
well-known among amateur astronomers, and an ideal object for public observations because of the very different
colors of its components. It has been observed for over more than two centuries \citep{hass16}.
However, there still exist many open questions about the origin, the components, and the orbital parameters of the 
Albireo system. 
Because of this, Albireo is still studied today and has been observed with many different methods 
\citep{scardia07,hartkopf1999,mason13,roberts18,IAU198}.
There is the long-lasting question as to whether the two bright components Albireo~A and B form a 
gravitationally bound system or not \citep{griffin99}. 
The {\it Gaia} collaboration \citep{gaiamission} published improved parallaxes in the recent Data Release EDR3 \citep{gaiaedr3,gaiaedr31}, 
and it seems that a gravitationally bound orbit is still possible because the parallaxes of the components A and B
lie within a $2\sigma$ error.
It is worth mentioning that {\it Gaia}, in general, has
difficulties determining parallaxes of bright stars such as Albireo \citep{drimmel19}.
In a thorough spectral analysis, \citet{drimmel21} have shown that Albireo A and B have about the same age and, 
therefore, might have at least a common origin. This conclusion is also supported by the discovery
of a moving group containing four other fainter stars.

The star Albireo~A is a binary system, which has been known for a long time \citep{maury97,clerke99}.
The spectrum is a composite of a K3II giant with a B9V companion \citep{markowitz69,parsons98}. 
This binary system has also recently been studied
so as to understand its nature and to determine its properties \citep{jack18,bastian18,drimmel21}.
\citet{drimmel21} determined the orbital parameters of the Albireo Aa, Ac system 
combining a large set of long-term observations of the radial velocity (RV), astrometry, and speckle interferometry. 
They determined the orbital period of the Albireo Aa, Ac system to be about 122~years. 
Because of this long period, the uncertainties of the orbital parameters are still quite large. 
They detected the problem that the total mass of the Albireo~A system was too high when
compared to the masses determined by spectroscopic analysis.
This "missing mass" must be hidden in the star Albireo~Ac.
They proposed that Albireo~Ac might be a binary
system either consisting of two very similar main sequence stars or that there is an invisible black hole component.
This indicates that there exists at least one further object in the Albireo system.

In addition, there have been reports about the detection of another close companion
of Albireo~Aa using speckle interferometry \citep{Bonneau80,prieur02}.
However, the detections were only marginal so that the existence of the star Albireo~Ab has not been confirmed 
and is still doubtful.

In conclusion, the number of objects in the Albireo system and their properties are still unclear.
In this publication, we present the discovery of a new low mass star in the Albireo system orbiting
Albireo~Aa, which we detected with RV measurements.

\section{Observations of Albireo A}

We obtained a time series of optical spectra with the Heidelberg Extended 
Optical Range Spectrograph (HEROS) echelle spectrograph mounted on the 
1.2~m Telescopio Internacional de Guanajuato Rob\'otico Espectrosc\'opico (TIGRE) 
telescope \citep{schmitt14} in central Mexico. 
The spectrograph has a resolution of $R\approx20,000$ and covers the 
optical wavelength range from about 3800 to 8800~\AA\ divided into two channels (red and blue) 
with a small gap of 120~\AA\ between the two channels at around 5800~\AA.
We observed Albireo~A for over three years.
All spectra were obtained with a high signal-to-noise ratio (S/N) between 200 and 400.
One goal of this study was to determine the RVs of
Albireo~Aa. We used the method described in detail in \citet{mittag18}.
This method has also been successfully
used in studies of spectroscopic binary stars \citep{jack20,jack21}. 

The observed spectra of Albireo~A are composite spectra that contain contributions
from the two stars Albireo~Aa and Ac (see \citet{drimmel21} for a complete TIGRE spectrum of Albireo~A).
The contribution from Albireo~Ac is only visible in the blue channel of HEROS 
because it is a B type main sequence star.
The main contribution in the red channel of the HEROS spectrograph comes from the K red giant
Albireo~Aa. The method to determine RVs uses only spectral lines in the red channel
so that all the lines correspond to Albireo~Aa.

\begin{table}
\caption{RV measurements of Albireo Aa in terms of the Julian date (JD) obtained with optical spectra using the TIGRE telescope.}
\label{tablerv}
\centering
\begin{tabular}{c c}
\hline\hline
JD & RV in km~s$^{-1}$ \\
\hline
2458391.55413 & $-24.83 \pm 0.13$ \\
2458446.55949 & $-25.07 \pm 0.12$ \\
2458527.01935 & $-25.48 \pm 0.11$ \\
2458528.01464 & $-25.56 \pm 0.11$ \\
2458529.01321 & $-25.52 \pm 0.10$ \\
2458577.96934 & $-25.58 \pm 0.10$ \\
2458612.92724 & $-25.47 \pm 0.10$ \\
2458624.88997 & $-25.38 \pm 0.11$ \\
2458717.65814 & $-24.83 \pm 0.13$ \\
2458748.56362 & $-24.88 \pm 0.13$ \\
2458770.59632 & $-24.95 \pm 0.13$ \\
2458807.53644 & $-24.89 \pm 0.12$ \\
2458898.01421 & $-25.34 \pm 0.10$ \\
2458933.92625 & $-25.44 \pm 0.10$ \\
2458962.96438 & $-25.49 \pm 0.10$ \\
2458991.86885 & $-25.50 \pm 0.10$ \\
2459021.80815 & $-25.38 \pm 0.10$ \\
2459053.76334 & $-25.06 \pm 0.11$ \\
2459054.79683 & $-24.97 \pm 0.10$ \\
2459084.75071 & $-24.75 \pm 0.14$ \\
2459116.59518 & $-24.90 \pm 0.12$ \\
2459145.57801 & $-24.91 \pm 0.12$ \\
2459155.55531 & $-24.89 \pm 0.13$ \\
2459165.53999 & $-24.82 \pm 0.12$ \\
2459175.57096 & $-24.88 \pm 0.13$ \\
2459186.54032 & $-24.79 \pm 0.13$ \\
2459257.02353 & $-25.33 \pm 0.11$ \\
2459259.01527 & $-25.24 \pm 0.11$ \\
2459269.01563 & $-25.43 \pm 0.11$ \\
2459270.01332 & $-25.38 \pm 0.11$ \\
2459292.94630 & $-25.42 \pm 0.11$ \\
2459303.92858 & $-25.40 \pm 0.10$ \\
2459313.90858 & $-25.44 \pm 0.10$ \\
2459323.87062 & $-25.41 \pm 0.10$ \\
2459333.96354 & $-25.36 \pm 0.10$ \\
2459416.81508 & $-25.03 \pm 0.10$ \\
2459434.75833 & $-24.84 \pm 0.11$ \\
2459446.61521 & $-24.73 \pm 0.14$ \\
2459495.57082 & $-24.81 \pm 0.13$ \\
2459524.54114 & $-24.84 \pm 0.12$ \\
2459536.53968 & $-24.88 \pm 0.12$ \\
2459546.53877 & $-24.85 \pm 0.12$ \\
2459562.54505 & $-24.76 \pm 0.15$ \\
2459629.01582 & $-25.27 \pm 0.10$ \\
\hline
\end{tabular}
\end{table}

\begin{figure}
\centering
\includegraphics[width=0.5\textwidth]{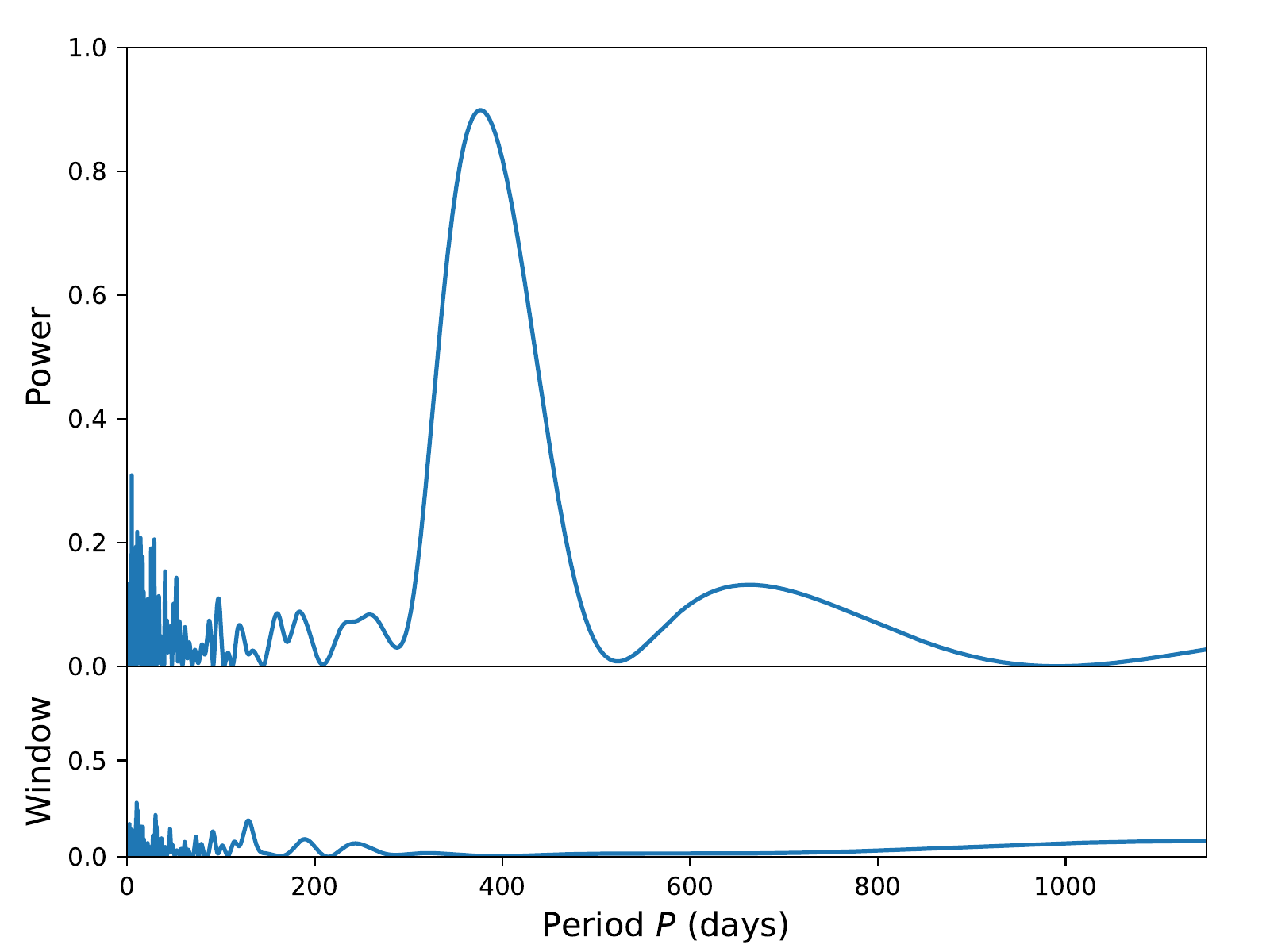}
\caption{Lomb-Scargle periodogram of the RV measurements of Albireo Aa showing a clear peak around 370 days.
The graph below shows the window function, where the peak disappeared.}
\label{fig:period}
\end{figure}
We observed Albireo~A for over three years and detected a clear variation in its RV.
We present all values of the RV measurements in Table~\ref{tablerv}, where we list the RV in km~s$^{-1}$
in terms of the Julian date (JD).
First, we checked for a possible periodicity in the RV measurements generating a Lomb-Scargle periodogram that
we demonstrate in Fig.~\ref{fig:period}. There is a clear and strong peak at around 370~days, which
indicates a periodic signal in the RV curve. The Fourier analysis gave a best period of 373.02 days.
Below, we plotted the window function, where the peak disappeared indicating that it is not just
a seasonal effect.

\begin{figure}
\centering
\includegraphics[width=0.5\textwidth]{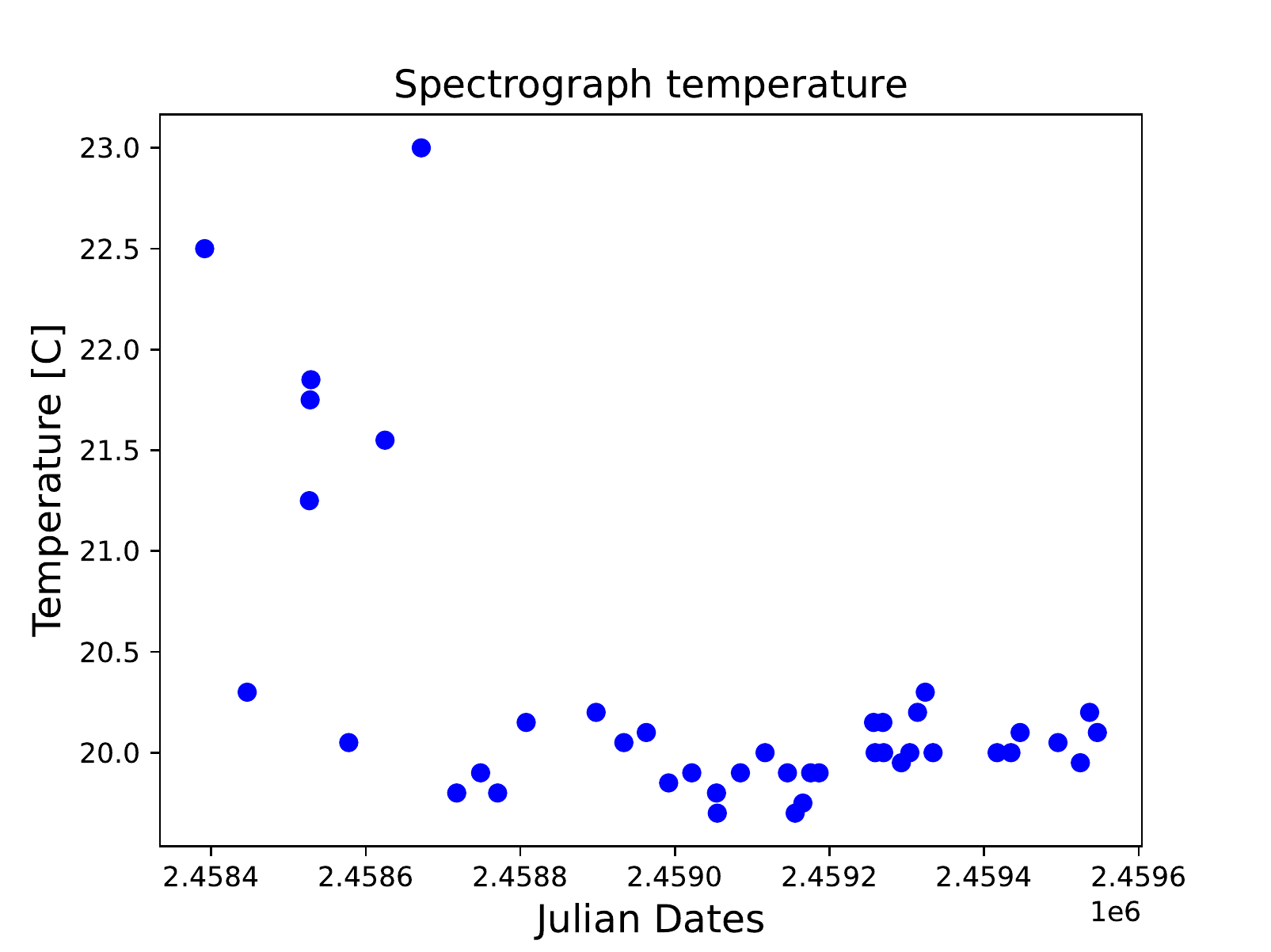}
\caption{Temperature of the spectrograph during the TIGRE observations of Albireo.}
\label{fig:spectrograph}
\end{figure}
With a periodicity so close to the terrestrial year, instrumental effects by seasonal influences
must rightfully be suspected, and the spectrograph temperature is the first issue that comes to 
mind. Nevertheless, the variations of the spectrograph temperature at each take of a spectrum 
(see Fig.~\ref{fig:spectrograph}) show no periodicity anywhere near a yearly cycle. Furthermore, it is important to note that the 
larger temperature variations in the first year of our observations was stopped by
the installation of a second stage air conditioning of the spectrograph and its encapsulation,
significantly improving the temperature stability for the latter years. There is no
noticeable change or scatter of the RV measurements between those in the first and 
those in the remaining periods observed. This is additional proof that the RV recordings
are real and not influenced by instrumental or seasonal effects.
Finally, we point out that in the same time interval, and with the same instrumentation and reduction procedures,
we measured RVs for about 50 other stars -- and none of them showed any periodicity close to one year.

\subsection{Stellar parameters of Albireo Aa}

We determined the stellar parameters of Albireo~Aa with our improved method with \texttt{iSpec} as
described in detail in \citet{rosas21}. We used the stellar parameters of \citet{drimmel21}
as input parameters and performed a set of 27 fit calculations using the
new method. This improved method more consistently determines the values for the surface gravity $\log{g}$, 
and for the rotational and turbulence velocities. 
These values are required for our analysis.

\begin{table}
\caption{Stellar parameters of Albireo Aa determined with an improved
method. Values of \citet{drimmel21} are shown for comparison.}
\label{tablestellar}
\centering
\begin{tabular}{ccc}
\hline\hline
Parameter & value & \citet{drimmel21}\\
\hline
$T_\mathrm{eff}$        & $4358\pm10$~K    & $4382.7\pm2.1$~K \\
$\log{g}$               & $1.68\pm0.03$    & $0.93\pm0.01$ \\
$[M/H]$                 & $-0.08\pm0.01$   & $0.02$\\
$[\alpha/Fe]$           & $-0.10\pm0.07$   & $0.08$\\
$v_\mathrm{mic}$        & $1.85\pm0.01$~km~s$^{-1}$ & $1.57$~km~s$^{-1}$\\
$v_\mathrm{mac}$        & $2.55\pm0.29$~km~s$^{-1}$ & $5.22$~km~s$^{-1}$\\
$v_\mathrm{rot}\sin{i}$ & $4.45\pm0.20$~km~s$^{-1}$ & $8.34\pm0.4$~km~s$^{-1}$\\
\hline
\end{tabular}
\end{table}

We present the results of our spectral analysis in Table~\ref{tablestellar}.
The effective temperature $T_\mathrm{eff}$ has hardly changed in comparison to \citet{drimmel21}.
The value for the surface gravity ($\log{g}=1.7$) is now consistent
with the value calculated based on the parallax. 
The metallicity $[M/H]$ is slightly lower.
The primary goal was to determine the rotational velocity, 
which has a value of $v_\mathrm{rot}\sin{i}=4.45$~km~s$^{-1}$. 
Using the value for $\log{g}$ and assuming a mass of $5.2\;M_\odot$, 
we obtained a radius of $55\;R_\odot$ for Albireo~Aa.

%
%
\section{The orbital parameters of Albireo Ad}

We analyzed the RV measurements of Albireo Aa using 
the Radial Velocity modeling toolkit \citep[RadVel]{radvel} (version 1.4.9)
that models RV data using the method of Bayesian inference.
This toolkit fits Keplerian orbits to observed RV curves and determines the six orbital parameters,
which are the orbital period $P$, the time of inferior conjunction $T_c$,
eccentricity $e$, semi-amplitude $K$, the argument of the periapsis
of the star's orbit $\omega$, and the RV of the system $v_\mathrm{rad}$.
To obtain an estimation of the uncertainties, RadVel contains the Markov-Chain Monte Carlo (MCMC)
package of \citet{foreman2013}.

Because Albireo~Aa is a member of a binary system with Albireo~Ac,
which has an orbital period of about 122~years \citep{drimmel21}, 
there is a slow change in the measured RV during the time of our observation campaign. 
The orbit determined by \citet{drimmel21} has large uncertainties, especially during the
period of our observations (2018 to 2021). 
In their Figure~6, the possible orbits may differ in up to 3~km~s$^{-1}$ in the RVs around the year 2020.
This is larger than the semi-amplitude of the orbit.
Therefore, we decided to fit the effect of an RV change induced by Albireo~Ac using the $dvdt$ parameter of RadVel, 
which fits a linear change to the observed RV curve.

\begin{table}
\caption{Orbital parameters of Albireo Ad}
\label{tablepara}
\centering
\begin{tabular}{c c}
\hline\hline
Parameter & value \\
\hline
$P$ & $371.2\pm5.6$~days \\
$K$ & $0.341\pm0.025$~km~s$^{-1}$\\
$e$ & $0.062\pm0.057$ \\
$\omega$ & --- \\
$T_c$ & $2458867\pm18$~JD\\
$v_\mathrm{rad}$ & $-25.158\pm0.019$~km~s$^{-1}$\\
$dvdt$ & $0.000172\pm0.000057$~km~s$^{-1}$~day$^{-1}$\\
\hline
\end{tabular}
\end{table}

In Table~\ref{tablepara}, we list the orbital parameters determined with RadVel.
The errors were obtained with an MCMC run using 100 walkers and 20000 steps.
The orbital period of the discovered companion Albireo~Ad is $P\approx371$~days. 
The semi-amplitude
with $K=0.341$~km~s$^{-1}$ is quite small, but still above the detection limit
of 0.1~km~s$^{-1}$ of HEROS/TIGRE \citep{mittag18}.
The orbit is within the error circular ($e=0.062\pm0.057$) and, therefore, we could not
determine the argument of the periapsis of the star's orbit $\omega$.
It is important to note that the determined RV of the system ($v_\mathrm{rad}=-25.158\pm0.019$~km~s$^{-1}$)
is not the systemic RV of the Albireo~Aa, Ad system 
because it includes the effect of Albireo~Ac.

Considering the orbit of Albireo~Ac with a period of 122 years and a
semi-amplitude of 2.91~km~s$^{-1}$, one obtains an
average change of 0.00026~km~s$^{-1}$ per day, 
which is consistent with the fitted value of $dvdt$.
The general trend of an increasing RV is also correct because the RV curve
of Albireo~Aa, Ac passed through the minimum a few years ago (around 2010).
It is important to keep in mind that the best fit RV curve of \citet{drimmel21} is not very
well defined.

\begin{figure*}
\centering
\includegraphics[width=0.90\textwidth]{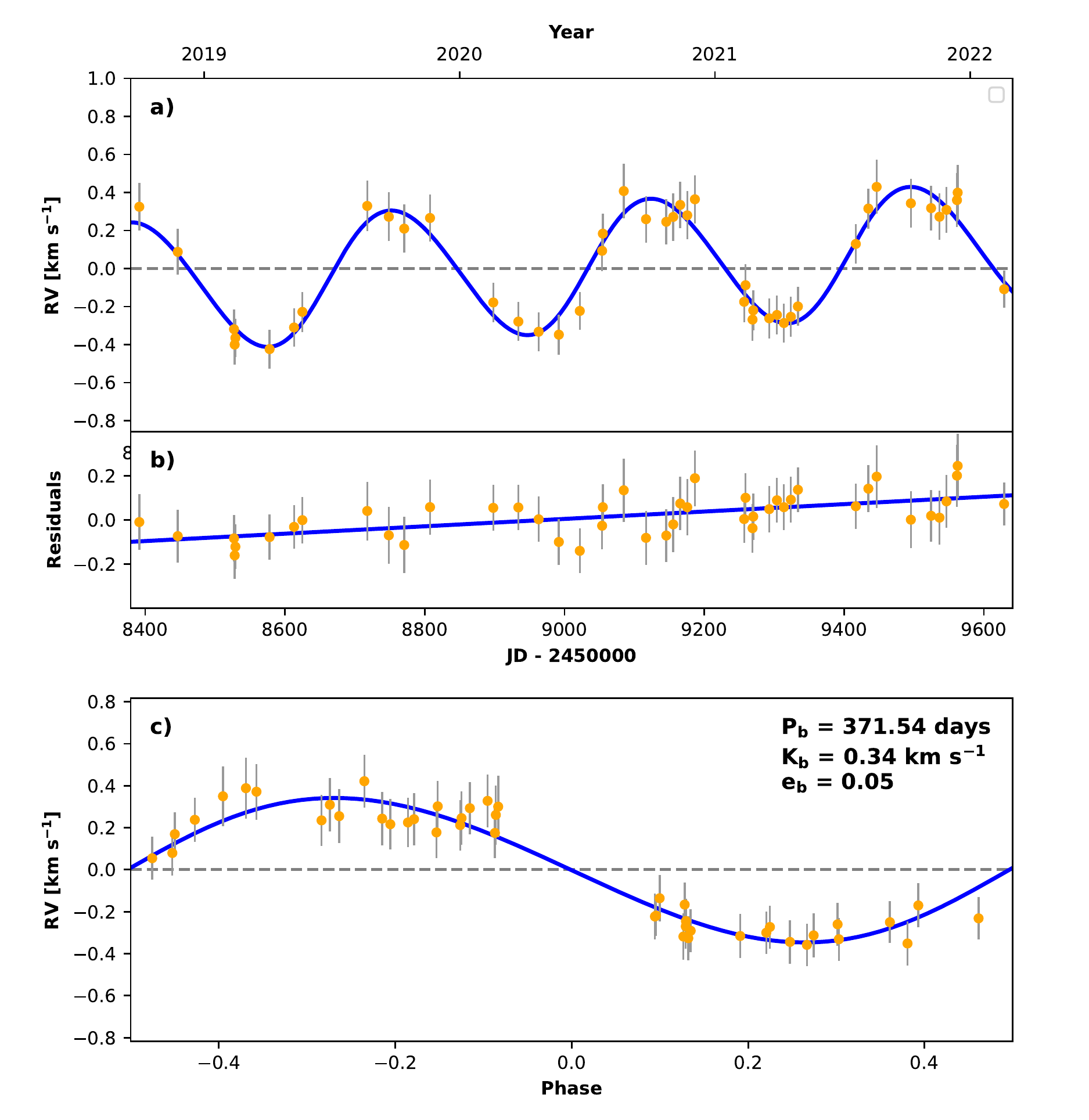}
\caption{RV curve of Albireo Aa with the fit curve obtained with RadVel.}
\label{fig:rv_curve}
\end{figure*}

We present the complete RV curve and the RadVel fit of Albireo~Aa in Fig.~\ref{fig:rv_curve}.
Subplot a shows the complete data set of the TIGRE RV measurements. The solid line represents
the RadVel fit. The residuals are shown in subplot b. Here, the solid line includes the linear trend of
the fitted $dvdt$ parameter.
The phase-folded RV curve is demonstrated in subplot c. The part around phase 0.0 has no RV measurements because
the period of the signal is very close to one year, and Albireo~Aa was not observable during that phase.

\begin{figure}
\centering
\includegraphics[width=0.5\textwidth]{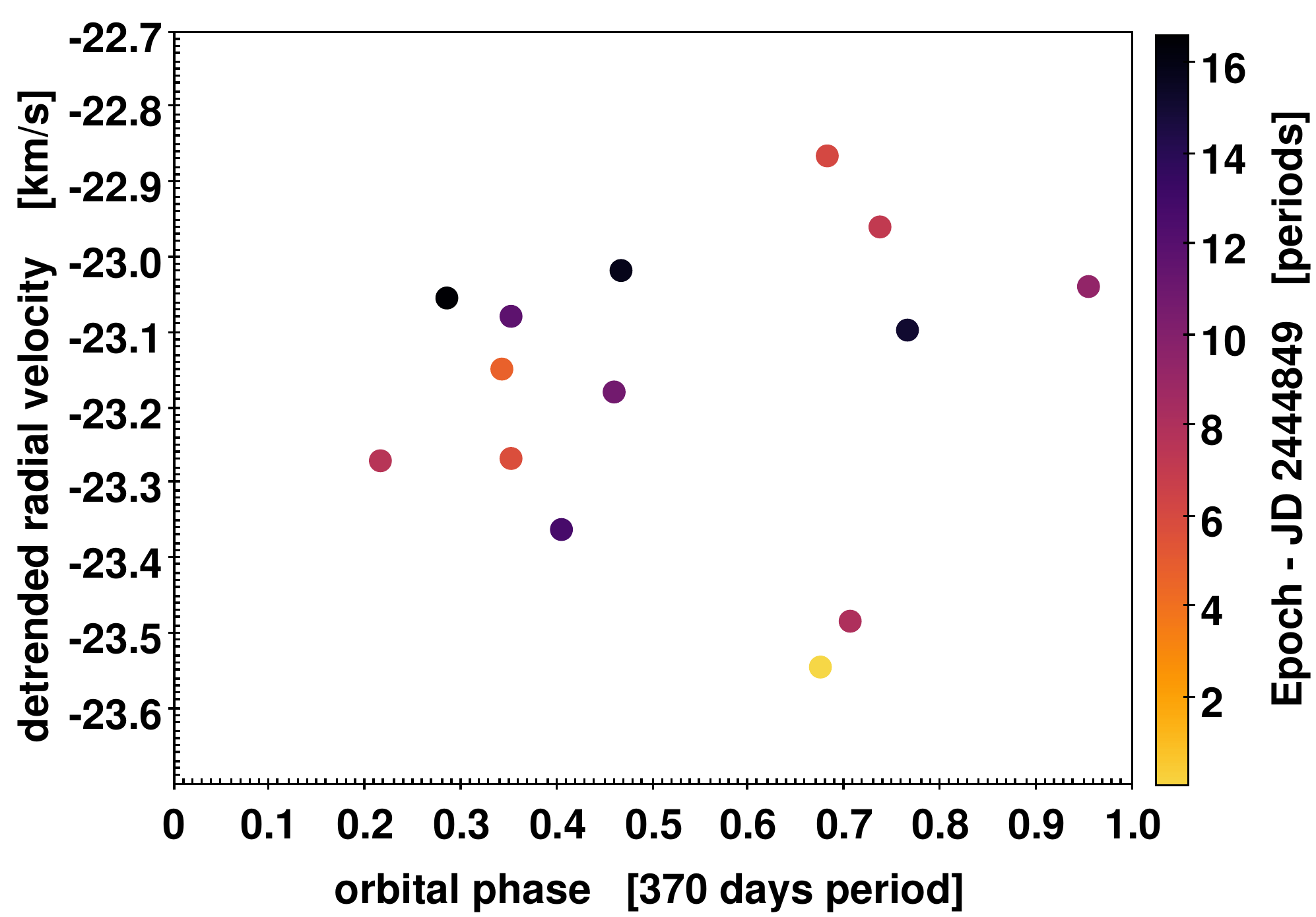}
\caption{RV measurements of CORAVEL. Vertical axis: CORAVELe RV after subtraction of a linear trend. 
Horizontal axis: Epochs phased with 370 days. Color: Epoch (in units of 370 days), to make any remaining trends visible.}
\label{fig:coravel}
\end{figure}
There exist published measurements of RVs of Albireo~Aa for over a decade in the literature \citep{drimmel21}.
The oldest measurements  obviously have a very large uncertainty. We took a closer look into the 14 RV measurements obtained 
with CORAVEL on the 1-m Swiss telescope at the Haute-Provence Observatory \citep{coravel}. 
Unfortunately, the CORAVEL RV data still have a scatter that is too large to detect any clear RV variation in the range of the
amplitude of Albireo~Ad.
In Fig. \ref{fig:coravel}, we present the CORAVEL data. We corrected the data for a linear trend caused by Albireo~Ac.
The color-coding according to the observation epoch clearly shows that there is no other trend in the data. Thus, we could
not use the CORAVEL data to further constrain the period of the Albireo~Ad orbit.

%
%
\section{The mass of Albireo Ad}

We calculated the mass function $f$ of the Albireo Aa, Ad system using the known formula
\begin{equation}
\label{eq1}
f=\frac{M_\mathrm{Ad}^3\sin^3{i}}{(M_\mathrm{Aa}+M_\mathrm{Ad})^2}=\frac{P K^3}{2\pi G}(1-e^2)^{3/2},
\end{equation}
where $G$ represents the gravitational constant, $M_\mathrm{Aa}$ and $M_\mathrm{Ad}$ are the
masses of Albireo Aa and Ad, respectively, and $i$ is the inclination of the orbit.
The value $f=0.00000152 \pm 0.00000036\;M_\odot$ is quite small, indicating
a small mass for the companion. 
We assumed a value of $M_\mathrm{Aa}=5.2\pm0.1\; M_\odot$ for the stellar mass of Albireo~Aa \citep{drimmel21}.
Solving the above equation, 
we calculated the minimal mass of Albireo~Ad to be $M_\mathrm{Ad}\sin{i}=0.0345\pm0.0026\;M_\odot$.
This corresponds to about $36$ Jupiter masses ($M_J$).
We assumed that the inclinations $i$ of the Albireo~Ac and the Albireo~Ad orbits are the same.
\citet{tokovinin17} found a strong tendency of orbit alignment in triple stars when the
members are within 50~AU.
With the inclination $i=156.15^\circ$ \citep{drimmel21},
the mass of Albireo~Ad is $M_\mathrm{Ad} = 0.085\pm0.007\; M_\odot$.
Using the Kepler's law equations, the maximal orbital separation of the two stars is about 1.9~AU.
Assuming a distance of approximately 120~pc to Albireo~A, the angular separation in the sky
is 0.0157~arcsec.

\citet{kervella19} found a signal in their analysis of proper motion anomaly in {\it Gaia} DR2 data. 
They determined a secondary mass of $m_2^\dagger=416.82^{+132.78}_{-77.84}M_J$.
This signal of an anomaly in the proper motion can be caused by both the
distant companion Albireo~Ac or the newly discovered close companion Albireo~Ad. 
Since the proper motion anomaly is actually a vector, 
the measured effect is a combination of the influences from both companion stars.
The relation between the \citet{kervella19} mass $m_2^\dagger$ and the actual mass $m_2$ is given by
$m_2^\dagger=m_2/\sqrt{r}$, with $r$ being the orbital radius. 
For Albireo~Ac, we have $r = 48$~AU and $m_2=2.7\;M_\odot=2828.44\;M_J$. 
The contribution from Albireo~Ac is, therefore, $m_2^\dagger= 408\;M_J$, which coincides
with the signal found by \citet{kervella19} very well.
We performed the same exercise for Albireo~Ad, with $r=1.9$~AU and $m_2=0.085M_\odot = 89\;M_J$.
The contribution from Albireo~Ad to the signal of a proper motion anomaly 
found by \citet{kervella19} is just $m_2^\dagger= 64.6\;M_J$.
In addition, as described in \citet{kervella19}, there is a smearing factor $\gamma$,
which in the case of Albireo~Ad is $\gamma=0.1$. Thus, the final observable signal is just
$6.5~M_J$.
This means that the signal found in \citet{kervella19} almost completely originates in
the influence from Albireo~Ac, and there is no contribution from Albireo~Ad.

%
%
\section{Discussion}

We first checked if our detected star is actually the already reported star Albireo~Ab and
found that this is not the case because the angular separations are too different.
We found a maximal angular separation of $0.0158$~arcsec for Albireo~Ad from Aa.
The angular separations reported for the detections of Albireo~Ab
are $0.125$~arcsec \citep{Bonneau80} and $0.045$~arcsec \citep{prieur02}. 
Even the smallest measurement is already three times larger than 
the maximal angular separation for Albireo~Ad.
Thus, our detected star cannot be Albireo~Ab.
Using Kepler's third law ($T^2\propto a^3$), the expected period of 
the Albireo~Ab orbit should be at least 5.2~years. 
We did not detect any signal of that period in our RV curve,
but the HEROS instrument is probably not sensitive enough.

The new \texttt{iSpec} method for the determination of stellar parameters
of \citet{rosas21} now gives more reliable values for the rotational velocities of stars.
The determined rotational velocities have a factor of $\sin{i}$
because of the inclination of the rotational axis, which is unknown.
Based on the rotational velocity determined with iSpec of $v_\mathrm{rot}=4.45$~km~s$^{-1}$, 
we calculated the rotation period assuming a radius of 55~$R_\odot$. 
The result is a rotation period of about 618~days. This is the 
maximum of the rotation period, 
and depending on the inclination the rotation period can be shorter.

Another way to check if we somehow detected the signal of stellar rotation
instead of the one of the orbit of a companion star is by stellar activity. 
We checked for periodic variations in the common activity indicators.
The TIGRE telescope has a data reduction pipeline that determines the S-index \citep{mittag16},
which is a measure for the magnetic activity of the \ion{Ca}{ii}~H\&K lines. 
We found a small peak ($\approx0.4$) in the Lomb-Scargle periodogram at 370~days.
However, this part of the spectrum is completely dominated by the flux of Albireo~Ac.
The routine that measures the S-index obviously corrects for the RV of Albireo~Aa. 
This introduces variations into the S-index, because
the flux of Ac is falsely shifted in wavelength and, therefore, in the window that is used
to determine the S-index.
Because the RV correction has the variation of the orbital period,
it obviously causes this false signal.

Variations of the RV in stars caused by stellar activity (star spots)
are a widely known problem for the detection of exoplanets. 
Several studies have been performed to quantify this effect \citep{saar97,desort07}.
As in the case of a companion, the signal is a sine curve, but 
there are usually jumps (phase shifts) in the signal because of the appearance and
disappearance of the star spots or active regions. We did not detect any
phase shifts during over three years of RV observations. 
The expected variations caused by stellar activity have normally peak-to-peak
amplitudes of at most 0.1~km~s$^{-1}$ \citep{saar97}. This is below our detected RV signal with 
a peak-to-peak amplitude of $0.68$~km~s$^{-1}$.
Thus, the detected signal of RV variations is caused by the orbital movement of a companion star
and cannot come from stellar activity on Albireo~Aa.

\begin{figure}
\centering
\includegraphics[width=0.5\textwidth]{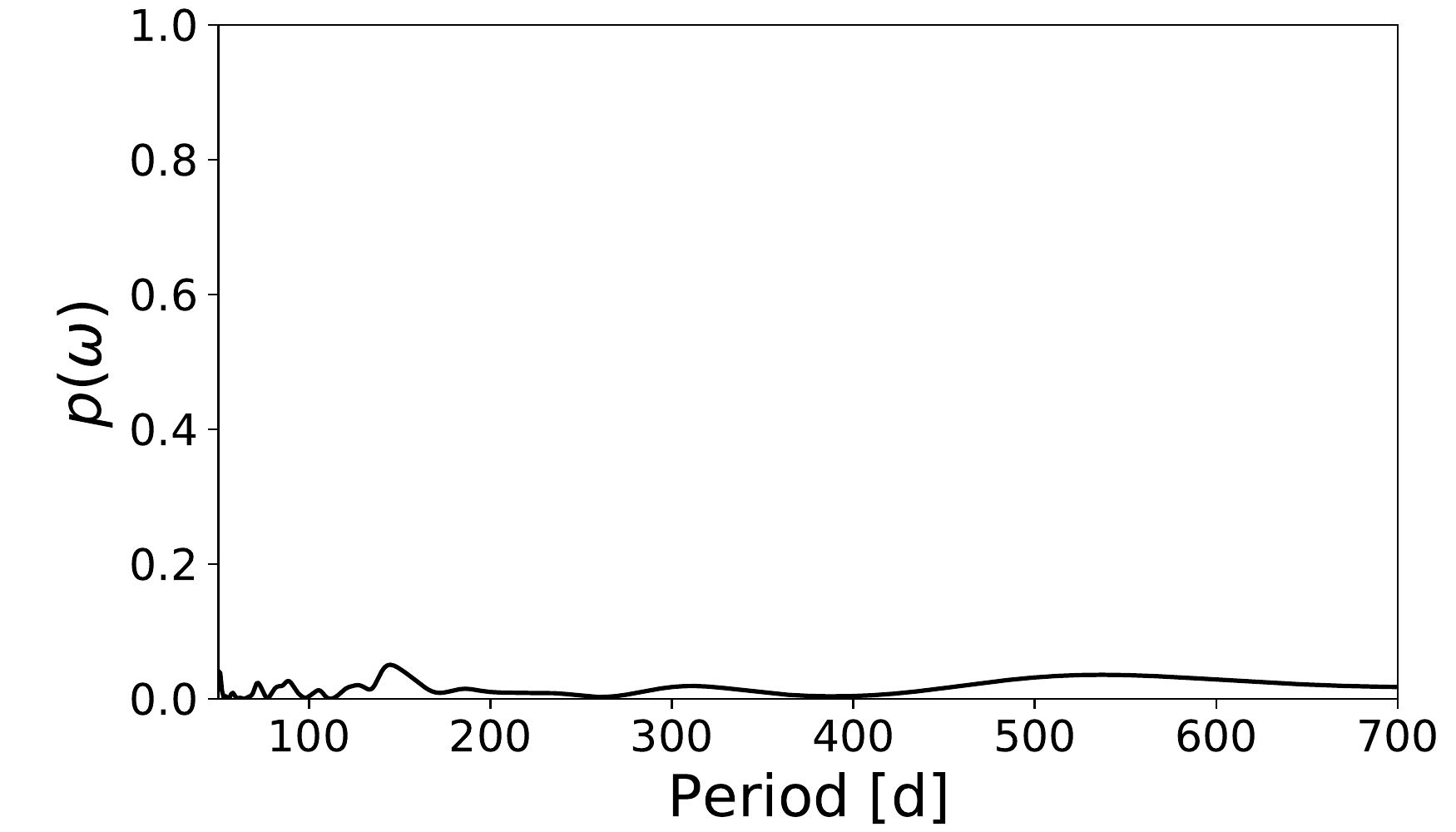}
\caption{Lomb-Scargle periodogram of the HIPPARCOS photometry of Albireo~A. There are no periodic variations.}
\label{fig:hipperiod}
\end{figure}

\begin{figure}
\centering
\includegraphics[width=0.5\textwidth]{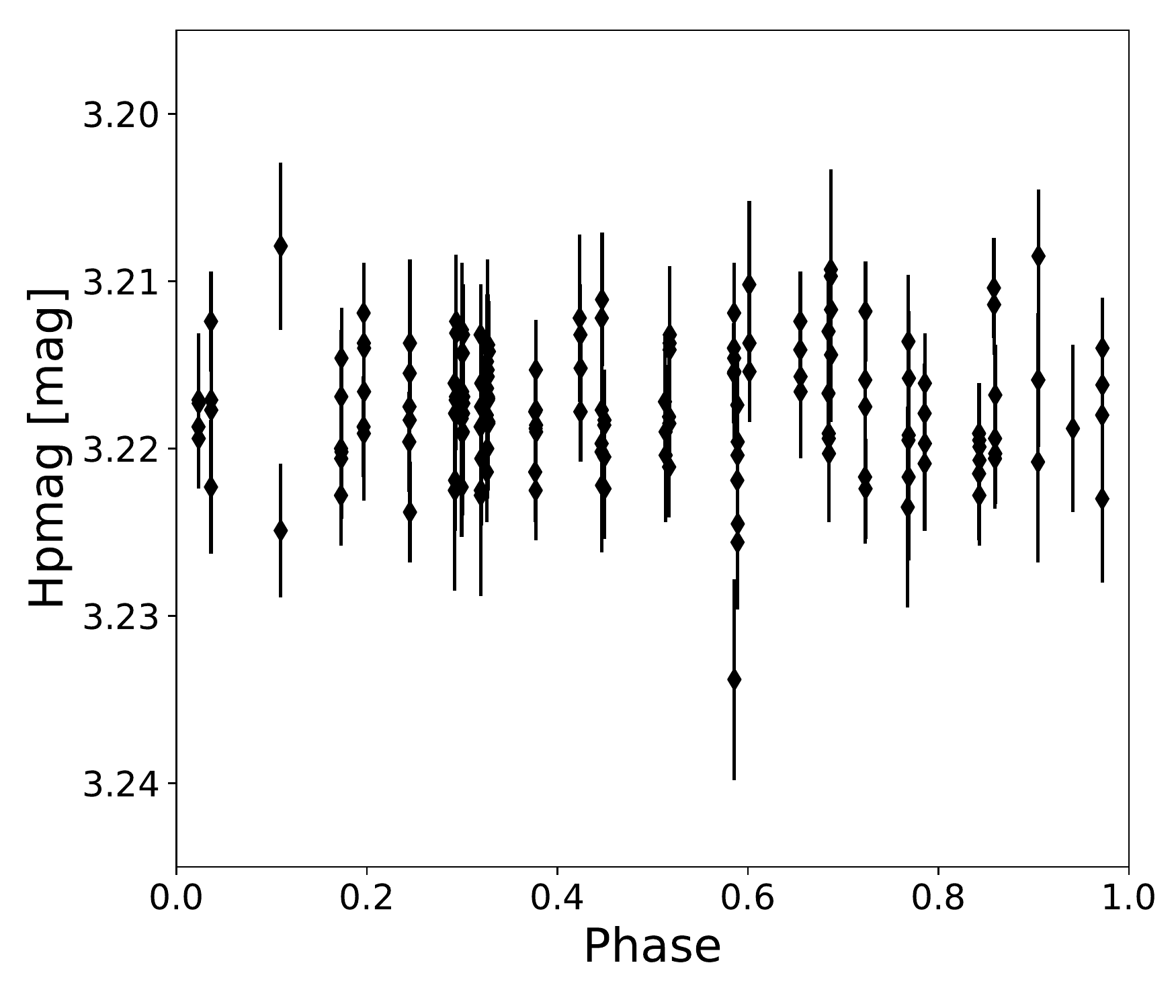}
\caption{HIPPARCOS photometry data of Albireo~A folded with the 371~days period show no variations.}
\label{fig:hipphase}
\end{figure}

We may comfortably exclude the possibility of a pulsation or oscillation for the RV signal.
The former possibility can be excluded by the absence of a matching photometric signal in the HIPPARCOS data \citep{hipparcos2}.
If there were any pulsations of Albireo Aa, then a brightness change should be observed.
As shown in Fig.~\ref{fig:hipperiod}, the periodogram of the HIPPARCOS photometry data has no peak at all. 
This can also be seen in Fig.~\ref{fig:hipphase} where the phase-folded photometry of Albireo~A is presented.
The idea of an oscillation is excluded by the expected period being much shorter
--- after all, Albireo Aa is a much more compact giant than a Mira star. According to \citet{kjeldsen95},
the frequency of the maximum-amplitude oscillation, $\nu_\mathrm{max}$,  of a star (which is the well-known
5-minute mode of the Sun)
mainly depends on gravity $g$ and effective temperature $T_\mathrm{eff}$ by being proportional to  $g/\sqrt{T_\mathrm{eff}}$. Using
the  physical parameters of Albireo Aa listed in Table~\ref{tablestellar},
$\nu_\mathrm{max}$ of Albireo Aa is exactly 500 times smaller than the 5-minute oscillation of the Sun,
meaning a period of 1.74 days.

The orbital period of 371~days is very close to one year. 
Because Albireo~A is at least a triple system, there must be a large difference
between the orbits of Ac and Ad, so that the system is gravitationally stable.
The existence of the star Albireo~Ab would actually disturb this configuration,
making the system unstable.

\citet{drimmel21} found evidence that Albireo~Ac itself is a binary system.
They determined the mass ratio of Albireo~Aa and Ac, 
and it indicates a "missing mass" for Albireo~Ac.
In this work, we discovered a new star in the Albireo~A system that
obviously changes the mass distribution.
However, the additional mass is very small (0.085~$M_\odot$) and has
no effect on the mass problem of the Albireo Aa, Ac system.

In passing, we note that -- depending on the unknown orientation of the Albireo~Ad orbit, 
that is to say depending on its projection onto the celestial sphere -- the 371-day period does modify
the astrometrically measured parallax of Albireo~A. 
If the inclination of the orbit is indeed about 156~degrees and if the semi-major axis is thus
about 16~mas, then the radius of the Aa orbit around the common center of mass is about 0.25~mas.
If the inclination is farther from 90 or 270~degrees, that is if the actual mass of Ad is higher
than 0.085\,$M_\odot$, the astrometric orbit of Aa is correspondingly larger. 
The parallaxes of Albireo A and B, as published in Gaia eDR3, are 8.98$\pm$0.45~mas and 8.19$\pm$0.08~mas, respectively. 
The difference of 0.79\,mas is a bit less than twice the combined uncertainty. 
The astrometric effect of the newly discovered component Ad might possibly
explain a significant part of this difference.

%
%
\section{Conclusions}

The famous bright multiple stellar system Albireo remains an object of interest,
to which it is worth dedicating further observations and studies.
In our analysis of the RV curve of Albireo~Aa, we detected a clear signal of a new member
of the system. Albireo~Ad is a star with a mass of 0.085~$M_\odot$ orbiting 
the giant star Albireo~Aa with a period of $P= 371$~days. The orbit is close to circular and
has a small semi-amplitude of $0.34~$km~s$^{-1}$.

We conclude that the Albireo system has a clear hierarchical structure.
There is the possible gravitational connection between Albireo~A and B. 
Albireo~A is a multiple stellar system. The companion Albireo~Ac
is probably a binary star \citep{drimmel21}.
In addition, the new member Albireo~Ad has a close orbit to Albireo~Aa.
With the discovery of the new star Albireo~Ad, the existence of Albireo~Ab is very unlikely.

\begin{acknowledgements}
This research has been made possible by the CONACyT-DFG bilateral grant No. 278156.
We thank the University of Guanajuato for the grants for the projects 036/2021 and 105/2021
of the {\it Convocatoria Institucional de Investigaci\'on Cient\'ifica 2021} and {\it 2022}.
This work has made use of data from the European Space Agency (ESA) mission
{\it Gaia} (\url{https://www.cosmos.esa.int/gaia}), processed by the {\it Gaia}
Data Processing and Analysis Consortium (DPAC,
\url{https://www.cosmos.esa.int/web/gaia/dpac/consortium}). Funding for the DPAC
has been provided by national institutions, in particular the institutions
participating in the {\it Gaia} Multilateral Agreement.  
This research has made use of the VizieR catalogue access tool, CDS, Strasbourg, France \citep{vizier}.
\end{acknowledgements}

\bibliographystyle{aa}
\bibliography{all}

\end{document}